%% file: main.tex
\documentclass[sigconf, screen]{acmart}

\PassOptionsToPackage{table,xcdraw}{xcolor}
\AtBeginDocument{%
  }

\copyrightyear{2026}
\acmYear{2026}
\setcopyright{rightsretained}
\acmConference[CHI EA '26]{Extended Abstracts of the 2026 CHI Conference on Human Factors in Computing Systems}{April 13--17, 2026}{Barcelona, Spain}
\acmBooktitle{Extended Abstracts of the 2026 CHI Conference on Human Factors in Computing Systems (CHI EA '26), April 13--17, 2026, Barcelona, Spain}\acmDOI{10.1145/3772363.3778806}
\acmISBN{979-8-4007-2281-3/2026/04}

\graphicspath{{figures/}{pictures/}{images/}{./}} 

\usepackage{amsmath}
\usepackage{algorithm}

\usepackage{color}
\usepackage{enumitem}
\usepackage{bbding}

\usepackage{enumitem}
\setlist{noitemsep,parsep=0pt,partopsep=0pt, leftmargin=10pt} 
\usepackage{listings}
\usepackage{multirow}
\usepackage{colortbl}
\usepackage{graphicx}

\input{setup.tex}
\begin{document}
\title[]{Legitimizing, Developing, and Sustaining Feminist HCI \\ in East Asia: Challenges and Opportunities}



\author{Runhua Zhang}
\authornote{Equally contribution, co-lead this meet-up.}
\email{runhua.zhang@connect.ust.hk}
\orcid{0000-0002-0519-5148}
\affiliation{%
  \institution{The Hong Kong University of Science and Technology}
  \city{Hong Kong SAR}
  \country{China}
}

\author{Ruyuan Wan}
\email{rjw6289@psu.edu}
\orcid{0000-0002-0357-5139}
\affiliation{%
  \institution{The Pennsylvania State University}
  \city{University Park, PA}
  \country{United States}
}
\authornotemark[1]

\author{Jiaqi (Ella) Li}
\email{li.jiaqi16@northeastern.edu}
\orcid{0000-0003-0758-4101}
\affiliation{%
  \institution{Northeastern University}
  \city{Boston, Massachusetts}
  \country{United States}
}

\author{Daye Kang}
\email{dk564@cornell.edu}
\orcid{0000-0003-0193-7228}
\affiliation{%
  \institution{Cornell University}
  \city{Ithaca, New York}
  \country{United States}
}

\author{Yigang Qin}
\email{yqin27@syr.edu}
\orcid{0000-0003-0193-7228}
\affiliation{%
  \institution{Syracuse University}
  \city{Syracuse, NY}
  \country{United States}
}

\author{Yijia Wang}
\email{wang.y.2fe2@m.isct.ac.jp}
\orcid{0009-0004-2250-9163}
\affiliation{%
  \institution{Institute of Science Tokyo}
  \city{Tokyo}
  \country{Japan}
}

\author{Ziqi Pan}
\email{zpanar@connect.ust.hk}
\orcid{0000-0002-5562-8685}
\affiliation{%
  \institution{The Hong Kong University of Science and Technology}
  \city{Hong Kong SAR}
  \country{China}
}

\author{Tiffany Knearem}
\email{tiffany.knearem@mbzuai.ac.ae}
\orcid{0000-0003-4928-6225}
\affiliation{%
  \institution{MBZUAI}
  \city{Abu Dhabi}
  \country{United Arab Emirates}
}

\author{Huamin Qu}
\orcid{0000-0002-3344-9694}
\affiliation{%
  \institution{The Hong Kong University of Science and Technology}
  \city{Hong Kong SAR}
  \country{China}  
}
\email{huamin@cse.ust.hk}

\author{Xiaojuan Ma}
\email{mxj@cse.ust.hk}
\orcid{0000-0002-9847-7784}
\affiliation{%
  \institution{The Hong Kong University of Science and Technology}
  \city{Hong Kong SAR}
  \country{China}
}

\renewcommand{\shortauthors}{Runhua Zhang, Ruyuan Wan, Jiaqi (Ella) Li, et~al.}

\input{sections/00-abstract}



\maketitle

\input{sections/01-motivation}

\input{sections/01-goals}
\input{sections/02-activities}
\input{sections/03-organizers}

\input{sections/04-community}

\bibliographystyle{ACM-Reference-Format}
\bibliography{main}
\end{document}

%% file: setup.tex
\definecolor{tablerowcolor}{rgb}{0.667,0.667,0.667 }
\definecolor{tablerowcolor2}{rgb}{0,0,0}
\definecolor{visual}{HTML}{e8efd9}
\definecolor{motion}{HTML}{fde7d5}
\definecolor{narrative}{HTML}{e2dce9}
\definecolor{audio}{HTML}{d6ebf2}
\definecolor{bluecrayola}{rgb}{0.12,0.46,1.0}

%% file: sections/00-abstract.tex
\begin{abstract}
Feminist HCI has been rapidly developing in East Asian contexts in recent years. The region's unique cultural and political backgrounds have contributed valuable, situated knowledge, revealing topics such as localized digital feminism practices, or women's complex navigation among social expectations. However, the very factors that ground these perspectives also create significant survival challenges for researchers in East Asia. These include a scarcity of dedicated funding, the stigma of being perceived as less valuable than productivity-oriented technologies, and the lack of senior researchers and established, resilient communities. Grounded in these challenges and our prior collective practices, we propose this meet-up with two focused goals: (1) to provide a legitimized channel for Feminist HCI researchers to connect and build community, and (2) to facilitate an action-oriented dialogue on how to legitimize, develop, and sustain Feminist HCI in the East Asian context. The website for this meet-up is: \url{https://feminist-hci.github.io/}.
\end{abstract}

%% file: sections/01-motivation.tex
\section{Background and Motivation}


Feminist HCI, coined by Shaowen Bardzell~\cite{bardzell2010a, bardzell2011b} in 2010, advocates for applying feminist theories and values to the design, critique, and evaluation of interactive technologies. For over a decade, this agenda has inspired a rich spectrum of studies, including the design of tools that empower women~\cite{pcos_women_health, feminism_care, sun2025, geng2025, park2025, henriques2025}, LGBTQ+ communities~\cite{cui2022, taylor2024, fiesler2016archive}, and other marginalized groups~\cite{ibrahim2024d, ibrahim2024c, oguamanam2023, petterson2024}, as well as the critique of technologies and interfaces~\cite{grimme2024, tuli2022, pcos_women_health, india_menstruation_taboo}. 
In the wake of this influential agenda, a growing number of scholars with East Asian backgrounds are expanding Feminist HCI research. Prior studies, for example, have examined localized forms of digital feminism practices~\cite{deng2024a, qin2024, wan2025a, pan2025, steen2023you}, women’s health and well-being as shaped by regional customs and taboos~\cite{sun2025, geng2025, mandai2025, wang2024j}, or equality of labor and work under regional economic and gendered norms~\cite{tang2022, zhao2024a, zhou2023}. These works contributed valuable, situated knowledge to the global community by highlighting the historical, political, and cultural dynamics unique to East Asia.


However, the very factors that ground these unique perspectives also create significant survival challenges for researchers in the region. Driven by pragmatism and utilitarian priorities~\cite{shen2005}, funding agencies often sideline Feminist HCI research, perceiving it as less ``technical'' or ``cutting-edge'' and less valuable than productivity-oriented technologies. At the same time, feminism is frequently framed as politically sensitive in East Asia~\cite{niranjana2007feminism, heng2013great}, and related scholarship such as Feminist HCI is often stigmatized, met with backlash, and at times even deemed ``inappropriate'' for serious funding~\cite{choi2024a}. The persistent marginalization has significantly constrained the field’s growth and legitimacy. Consequently, local senior Feminist HCI researchers and established communities remain scarce in East Asia, leaving junior scholars with limited guidance and few clear career paths to follow.

Yet, even in the face of these challenges, we are witnessing a vibrant surge of interest in Feminist HCI within East Asian contexts~\cite{wan2025a, deng2024a, qin2024, sun2025, tang2022, mandai2025}. In addition to the increasing number of relevant publications, this momentum has also been reflected in the organizers’ community-building practices. At CHI 2025, for example, several junior co-organizers first connected through shared interests. Through personal networks, we took the initiative to launch an informal Feminist HCI Interest Group among Asian researchers. This group attracted around 30 participants to a Feminist HCI lunch gathering in Yokohama, Japan. The initiative has since evolved into an ongoing Feminist HCI Reading Seminar Group, which now includes 102 members. 



%% file: sections/01-goals.tex
Motivated by the momentum of Feminist HCI in East Asia and our prior practices, we propose this meet-up with two main goals. (1) To provide a legitimized and efficient channel for Feminist HCI researchers, especially junior scholars, to connect and build community beyond the limits of individual networks. (2) Grounded in the challenges faced by Feminist HCI researchers in East Asia, to facilitate an action-oriented dialogue among researchers at different career stages around the urgent question: ``\textbf{\textit{How can we legitimize, develop, and sustain Feminist HCI in the East Asian context?}}'' With these two goals, we aim to connect and support researchers in working on solutions that enhance visibility, secure allies and funding, and foster mutual support. Given the established influence of Feminist HCI within CHI, we envision this meet-up as an opportunity to consolidate Feminist HCI in East Asia and to attract broader attention to the region.








%% file: sections/02-activities.tex
\section{Activities}

\textbf{Welcome \& Opening (10 min.)} 
Organizers will introduce the goals and agenda of the meet-up. Prof. Shaowen Bardzell will share brief remarks as a pioneering researcher in the field. 

\noindent\textbf{Activity 1: Pick a Word, Share a Story (30 min.)}
Participants will pick from prepared feminist HCI keywords (detailed in supplementary), introduce themselves,
and briefly explain why they chose the keyword and their connection to feminist HCI. This activity aims to create an inclusive starting point for connect and conversation.

\noindent\textbf{Activity 2: Navigating Our Feminist HCI Journeys (45 min.)}
Participants will be divided into small groups of 3–6 based on career stage or background, each facilitated by a co-organizer. Within these groups, participants will use a shared journey board and sticky notes (detailed in supplementary) to reflect on their experiences of navigating feminist HCI research, including shaping career paths, seeking mentors or mentees, securing grants, and promoting Feminist HCI. Afterwards, groups will reconvene to present key insights to the whole room, enabling broader exchange and collective reflection. 

\noindent\textbf{Closing \& Next Steps (5 min.)}
Organizers will synthesize the key takeaways and share follow-up channels. This includes collecting feedback on the meet-up and expectations for future gatherings through a short survey, as well as announcing the website where upcoming events, such as the Feminist HCI Reading Seminars, will be posted.

\noindent\textbf{Considerations for Creating a Safe(er) Space.} Given that feminism and Feminist HCI can be sensitive or contested topics, we will take steps to foster a safe(er) and respectful environment. Before the meet-up, we will draft and publish a concise Code of Conduct on our website to clearly communicate expectations for engagement. At the start of the meet-up, organizers will introduce shared community agreements and clarify facilitation roles. Facilitators will be prepared to gently mediate tense moments and support participants who experience discomfort. These measures aim to uphold feminist values of care and ensure inclusive conversations.


%% file: sections/03-organizers.tex
\section{Organizers}

%

\textbf{Runhua Zhang} is a Ph.D. student at The Hong Kong University of Science and Technology. Her research in Feminist HCI focuses on gender bias in AI, digital feminism~\cite{pan2025}, and women’s health~\cite{sun2025}. As an advocate, she initiated informal Feminist HCI meet-ups at CHI 2025 and currently organizes a Feminist HCI Reading Seminar with 102 members.  

\textbf{Ruyuan Wan} is a Ph.D. student at The Pennsylvania State University. Her work examines digital feminism on social media. Her CHI 2025 paper on women’s hashtag re-appropriation for audience control on Rednote received a Best Paper Honorable Mention~\cite{wan2025a}. She also co-organized the informal Queer in AI meet-ups at CHI 2024 and 2025. 

\textbf{Jiaqi (Ella) Li} is a Ph.D. student at Northeastern University. Her research focuses on the (re)design of sociotechnical systems for queer people and women. She also co-facilitated the informal Queer in AI meet-up at CHI 2025. 

\textbf{Daye Kang} is a Ph.D. candidate at Cornell University. She explores Human-AI collaboration and women's health. Her CHI 2025 paper on tracking and managing PCOS (polycystic ovary syndrome) received the Best Paper Award~\cite{pcos_women_health}. 

\textbf{Yigang Qin} is a Ph.D. student at Syracuse University. He studies how digital technologies configure labor and work, and his CSCW 2024 paper on gender-blindness and feminist activism received the DEI Recognition Award~\cite{qin2024}. 

\textbf{Yijia Wang} is a Ph.D. student at Institute of Science Tokyo. Her research examining how East Asian cultural concepts such as kawaii (Japanese “cute”) shape gender bias or stereotypes in technologies~\cite{wang2024j}. 

\textbf{Ziqi Pan} is a Ph.D. candidate at The Hong Kong University of Science and Technology. She studies human-human interaction through daily technologies. Her work explores the well-being of queer and women's online communities~\cite{pan2025}.

\textbf{Tiffany Knearem} is an Affiliated Assistant Professor at the Mohamed bin Zayed University of Artificial Intelligence (MBZUAI), with research interests in human-AI alignment. She co-organized the CHI 2024 and CHI 2025 Computational UI workshops, and the CHI 2025 Bi-Align SIG.

\textbf{Huamin Qu} is a Chair Professor and the founding Dean of the Academy of Interdisciplinary Studies (AIS) at The Hong Kong University of Science and Technology (HKUST). His research has been widely recognized at venues such as CHI and UIST, earning multiple Best Paper Honorable Mention Awards. 

\textbf{Xiaojuan Ma} is an Associate Professor at The Hong Kong University of Science and Technology (HKUST). Her work has received Best Paper Awards and Honorable Mentions at CHI and CSCW. She is serving in the organizing committees of CHI 2026, CSCW 2025, 2026 and program committees in MobileHCI , SIGGRAPH Asia 2024, etc. 


%% file: sections/04-community.tex
\section{Community of Interest}
While this meet-up topics are primarily focused on Feminist HCI research in East Asian contexts, it is open to all CHI 2026 attendees by nature. Although research in this area has been steadily increasing, the visibility of East Asian scholars and research within the global CHI community remains limited~\cite{gamage2025, lu2024b}. This event provides an important opportunity to bring together both scholars from the region and those with broader interests. By fostering discussion on unique challenges such as funding constraints and regional socio-political dynamics, our meet-up seeks to generate and share knowledge on how these challenges can be addressed from both regional and global perspectives. We believe such dialogues align closely with and will contribute to CHI’s core values such as diversity and inclusivity.